\documentstyle[bo99,psfig]{article}
\def\pn{\par\noindent}

\title{Multiple BeppoSAX Observations of IC 4329A to Probe 
the Origin of the Compton Reflection Component in Seyfert 1 Galaxies}

\author{M. Cappi$^{1}$, G. Di Cocco$^1$, M. Dadina$^1$, G. Malaguti$^1$,
M. Matsuoka$^2$, \\
G. Matt$^3$, G.C. Perola$^3$, L. Piro$^4$}

\affil{1) ITESRE-CNR, I-40129, Bologna, Italy \\
2) SURP-NASDA, Tsukuba, Ibaraki 305-3805, Japan \\
3) Dipartimento di Fisica, Universit\`a  degli Studi ``Roma Tre'',
I-00146, Roma, Italy \\
4) Istituto di Astrofisica Spaziale, IAS-CNR, I-00133, Roma, Italy}

\begin{document}

\maketitle

\begin{abstract}

IC 4329A is the brightest known Seyfert galaxy in hard ($\sim$ 2-30
keV) X-rays and is likely to be representative of Seyfert 1 galaxies
as a class. A recent 100 ks {\it BeppoSAX} observation (Perola et al. 1999)
clearly confirmed the presence of a warm absorber, a reflection
component (R $\sim$ 0.6), and a high-energy cut-off in the power 
law at E$_c$ $\sim$ 270 keV. Its richness in spectral features, 
combined with its large flux ($\sim$ 1.6 $\times$ 10$^{-10}$ erg cm$^{-2}$ s$^{-1}$ 
between 2-10 keV), 
make this target ideal for multiple observations
(in particular with {\it BeppoSAX}) to search for spectral variations.
Results obtained from 3 follow-up observations (40 ks each) are
presented here. The first and most important goal of this study was to 
probe the origin of the Compton reflection component observed in
Seyfert galaxies by monitoring the variability of the
reflection continuum and Fe K$\alpha$ line in response to primary continuum
variations. The second goal was to search for variability in the high 
energy cutoff. We obtain however no conclusive results on any of
these issues. In fact, all four observations unfortunately caught the 
source at almost the same flux, showing only little, and marginal, spectral changes
between different observations.

\end{abstract}

\section{Introduction}

{\it Ginga}, {\it ASCA} and recent {\it BeppoSAX} ~X-ray observations
of Seyfert 1 galaxies strongly support a general {\it observational} 
framework that includes: ionized absorption by a warm absorber (WA), a steep 
intrinsic power-law continuum, a Compton reflection hump and
associated neutral FeK${\alpha}$ line (narrow and/or broad), and a 
high-energy cutoff ($E_{\rm cutoff}$).
These observational results have raised the following questions:

\begin{itemize}

\item What is the location of the X-ray reprocessor? Is it the 
accretion disk, the broad-line region (BLR) and/or the molecular torus? 
Is the answer the same for all Seyfert 1s? 

\item Where is the location of the WA? Is it within the BLR and/or
the molecular torus or is it external to them? Is there only one WA?

\item Does $E_{\rm cutoff}$ differ from source-to-source (as currently
shown by {\it BeppoSAX}, Piro 1999, Matt 2000) or is it variable with time
and/or flux in any single object?

\end{itemize}

In order to address these questions, we have performed multiple (4)
observations of the prototypical Seyfert 1 galaxy IC4329A to 
obtain time-resolved spectral constrains on the WA, reflection 
continuum, Fe K$_\alpha$ line and, possibly, $E_{\rm cutoff}$.
This is potentially the best way (see also Weaver 2000) to establish whether these 
observational features are produced from material near the 
source (e.g. the accretion disk, in which case short lags, $<$1000 s, 
are expected between the reflection and continuum variations) 
or farther away (e.g. in the BLR or 
torus, in which case longer lags of at least $\sim$ 10$^{6-7}$ s 
are expected).

\section{The Multiple {\it BeppoSAX} Observations}

IC 4329A $(z=0.014)$ is a Seyfert 1 galaxy well studied in X-rays
(Miyoshi et al. 1988; Piro, Yamauchi \& Matsuoka 1990; Fiore et al. 1992;
Cappi et al. 1996) and hard X-rays (Fabian et al. 1993; Madejski et al. 1995;
Zdziarski et al. 1995a). It offers the best opportunity to address the
above mentioned variability issues because it is among the brightest
known Seyferts in X-rays (F(2-10keV)$\sim$1.6$\times$10$^{-10}$erg 
cm$^{-2}$s$^{-1}$) and shows only {\it moderate} variability 
on short ($<$ day) timescales, but large variability on longer timescales 
(a factor 2-3 in a few days), both below and above 10 keV (Done, 
Madejski \& Zycki 2000; Fabian et al. 1993).

{\it BeppoSAX} observed IC 4329A in 1996 for 100 ks (Perola et
al. 1999), and then subsequently 3 times in 1998, with the 3 observations
spaced approximately 5 days apart. As expected, the
source did {\it not} vary much {\it within} each
observation. Unfortunately, it did not vary much {\it between} the 4 
observations either (see Fig. 1), despite the large and systematic flux 
variations typically observed in IC4329A by RXTE (Done, Madejski \& Zycki 
2000).

Spectral ratios were produced between all observations, in search of any
model-independent spectral variations (Fig. 2). We find marginal
evidence for variability of the spectral features at $E<2$ keV and
at $E=8$--10~keV, but we find no clear variations in the
FeK$_\alpha$ and reflection component intensities, as well as in the
high-$E_{\rm cutoff}$. The variations at low energies are most likely due
to variability of the ionization state of the WA responding to
continuum variations, but complications due to the presence of a
possible scattered component and/or intrinsic soft (excess) component
cannot be ruled out.

Spectral fitting of each observation with a complex model including 
2 absorption edges, a steep power-law continuum, a reflection
component with associated FeK$\alpha$, and a high-$E_{\rm cutoff}$ gave
best-fit parameters similar to those reported by Perola et
al. (1999) with somewhat larger errors (e.g. 
$\Gamma \sim 1.85 \pm$ 0.1, R $\sim 0.6 \pm 0.2$, $E_{\rm cutoff}$ $\sim$
250 $\pm$ 100 keV). Fig. 3 shows the marginal 
variations of the FeK line intensity and equivalent width. Fig. 4
shows the (similar) best-fit broad-band spectra of two different observations.

The most interesting result of the present analysis is that 
despite a 30\% flux increase between obs. 1 and obs. 2, the line
intensity became surprisingly weak (compare Fig. 1 and Fig. 3). 
During the higher flux obs. 3 the line re-established its typical
value of $\sim$ 100-150 eV, though. 
In other words, there is some evidence (at a $\sim$ 2 $\sigma$
significance level) that the Fe K$\alpha$ line {\it does not} follow
``instantaneously'' (at least within 40 ks) the continuum variations,
but it {\it does} on a timescale between 4--10 days. 

\section{Conclusions}

\begin{itemize}

\item
Moderate variations of the absorption structure, most likely related
to the warm absorber, are apparently detected but require further investigation
for more quantitative conclusions.

\item
Despite the unprecedented statistics of these observations, we find
{\it no} significant spectral variations of the Compton reflection continuum, the
power law spectral index and the high-enegy cutoff. This is probably
because {\it BeppoSAX} unfortunately caught IC 4329A at a not too
different flux level (30\% variation) in all four observations 
hampering, thereby, any spectral variability.

\item
There is marginal evidence that the FeK$\alpha$ line is being
produced by the outer regions of the accretion disk in IC 4329A, thereby
suggesting a possible explanation for the lack of a clear broad FeK$\alpha$ line in this
object. 

\end{itemize}

\vspace{-0.5cm}
\begin{acknowledgements}
M.C. wishes to thank Martin Elvis and Fabrizio Nicastro for
usefull discussions. This analysis has made use of data prepared by the 
{\it BeppoSAX} Scientific Data Center.
\end{acknowledgements}

\vfill\eject

\begin{figure}
\parbox{6truecm}
{\psfig{file=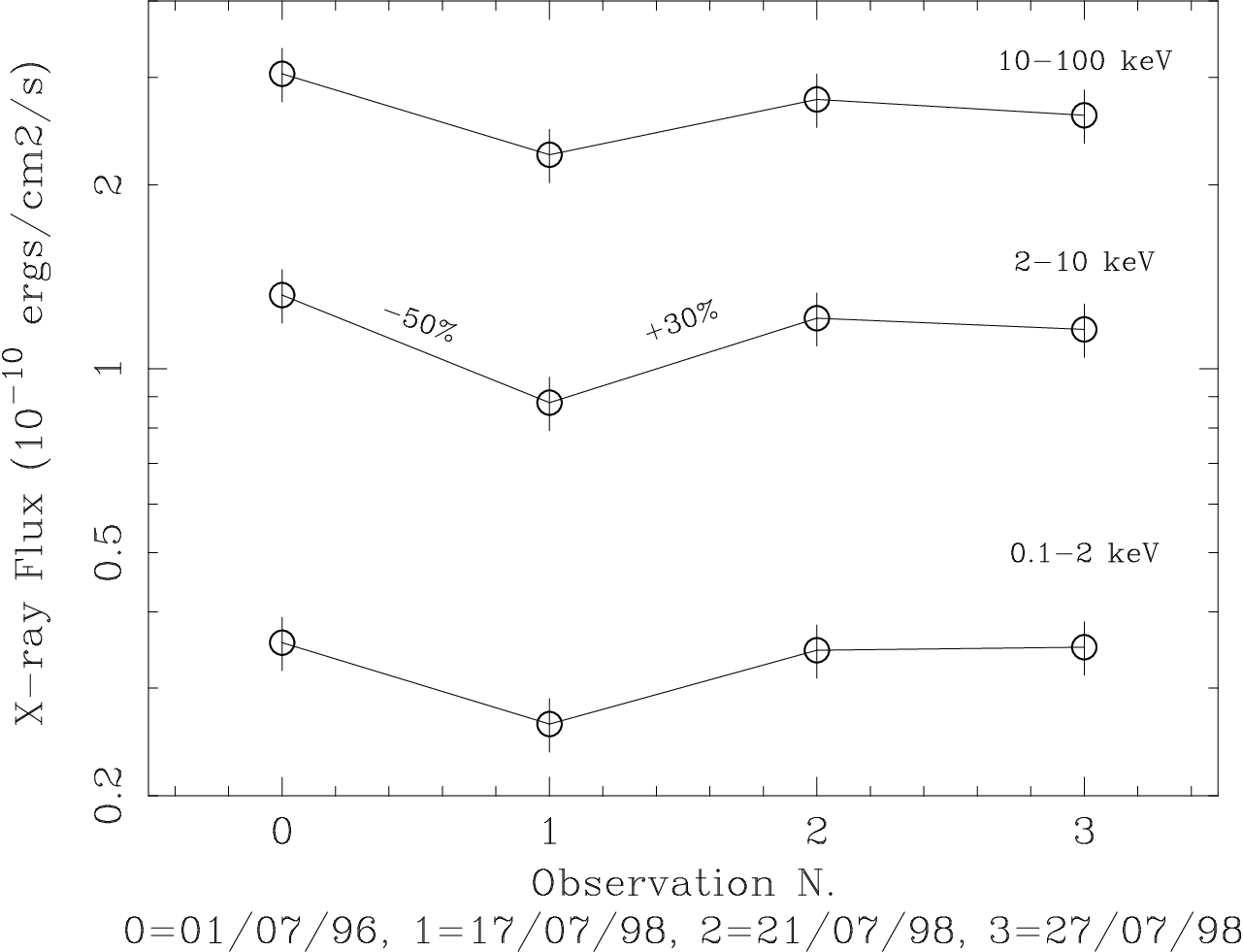,width=6cm,height=7cm,angle=-90}
\caption[h]{\small 0.1-2 keV, 2-10 keV and 10-100 keV flux of IC 4329A during Obs. 0
(the long 100 ks Obs.), Obs1, 2 and 3.}}
\parbox{6truecm}
{\psfig{file=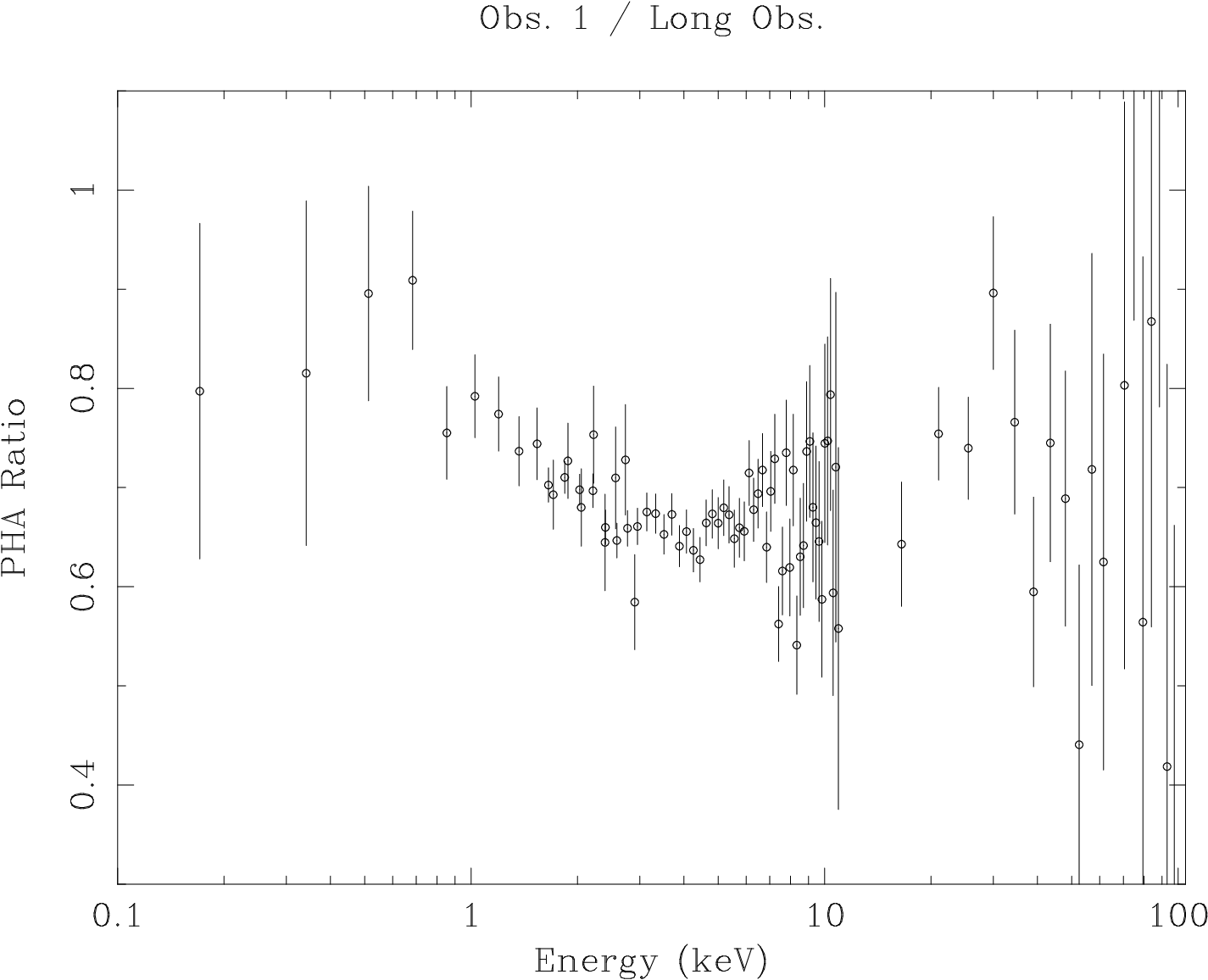,width=6cm,height=2.3cm,angle=-90}
\psfig{file=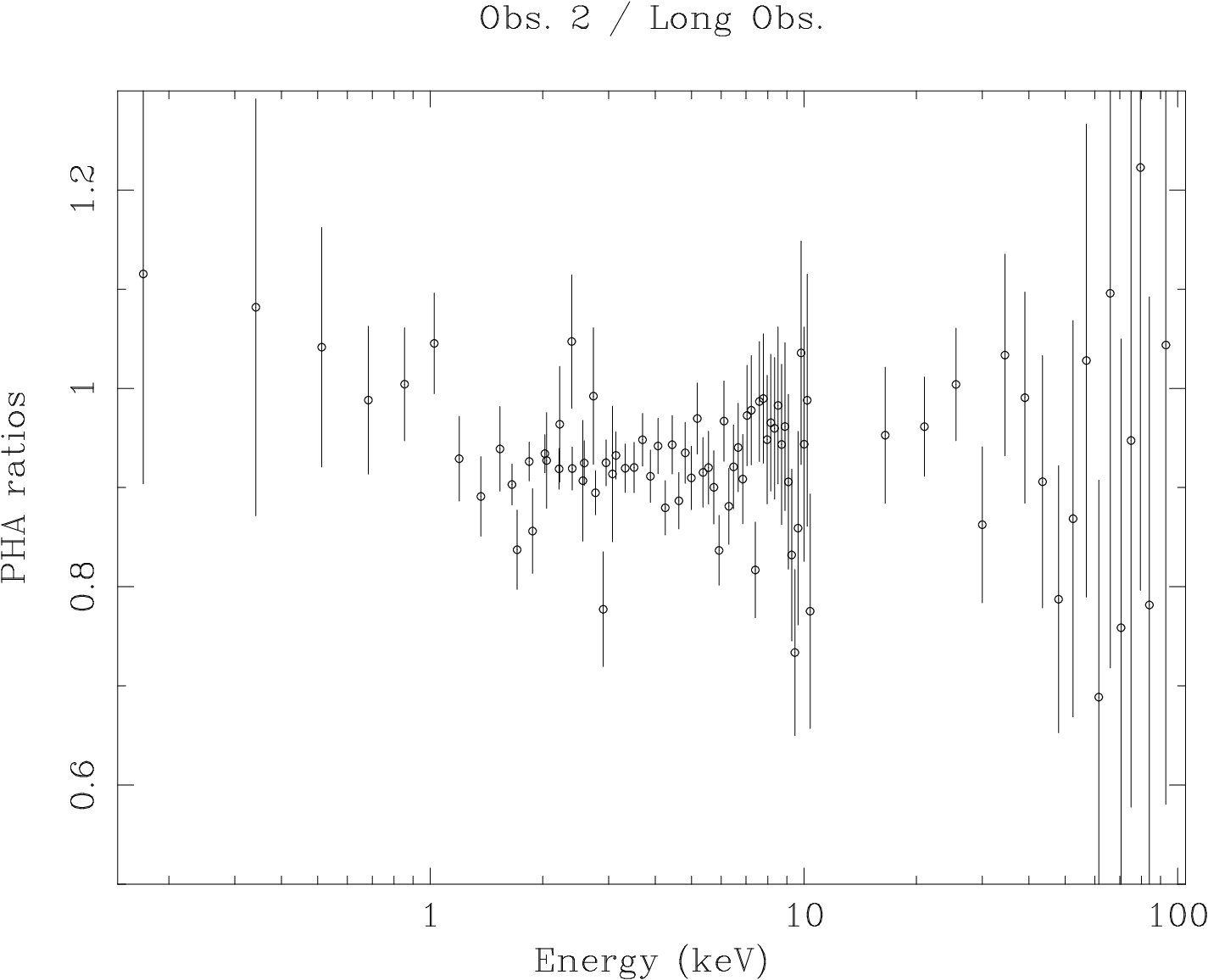,width=6cm,height=2.3cm,angle=-90}
\psfig{file=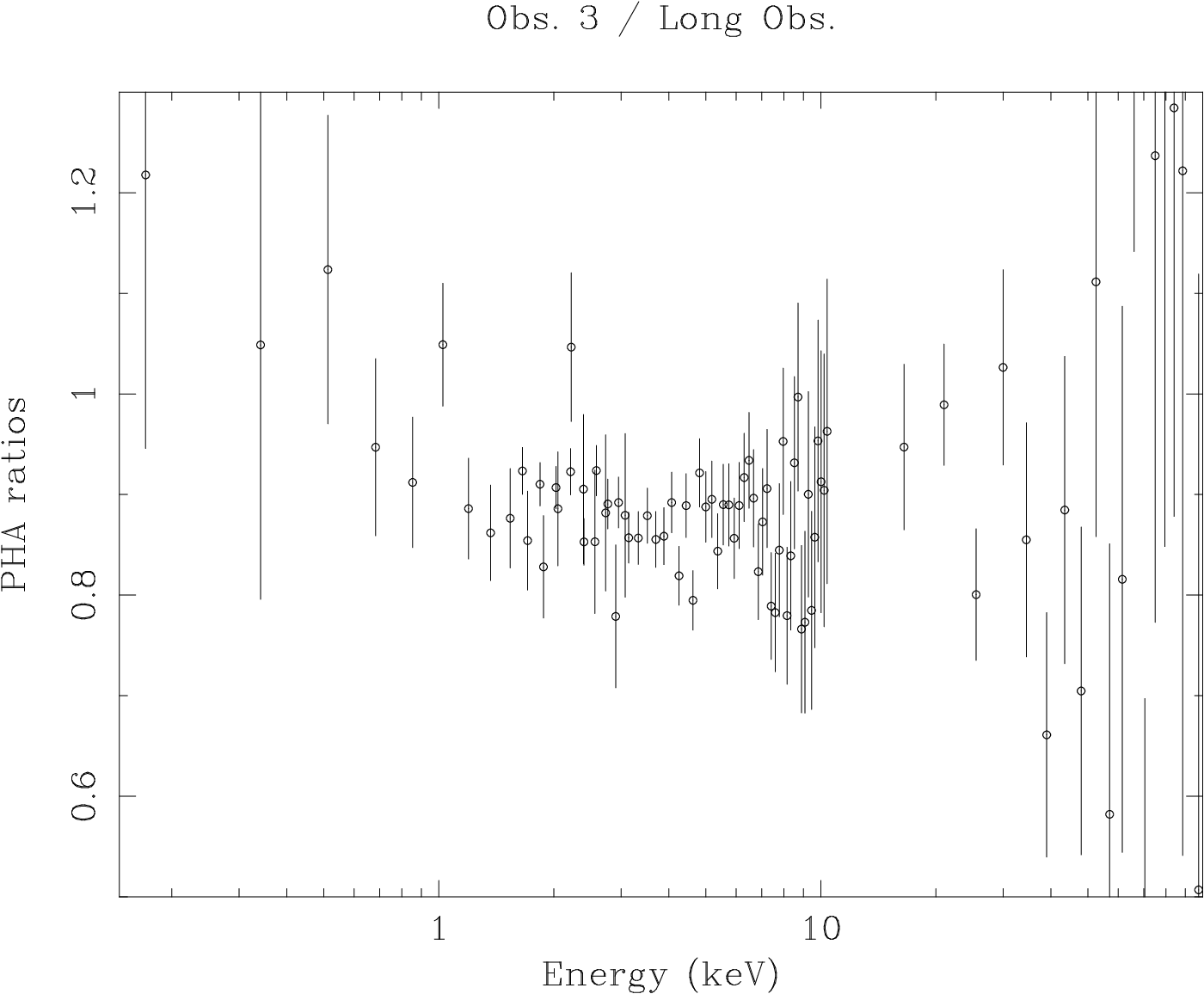,width=6cm,height=2.3cm,angle=-90}
\caption[h]{\small Spectral ratios of ${Obs.1}\over{LongObs}$ (top pannel),
${Obs.2}\over{LongObs}$ (mid pannel) and ${Obs.3}\over{LongObs}$ (lower pannel)}}
\end{figure}

\begin{figure}
\parbox{6truecm}
{\psfig{file=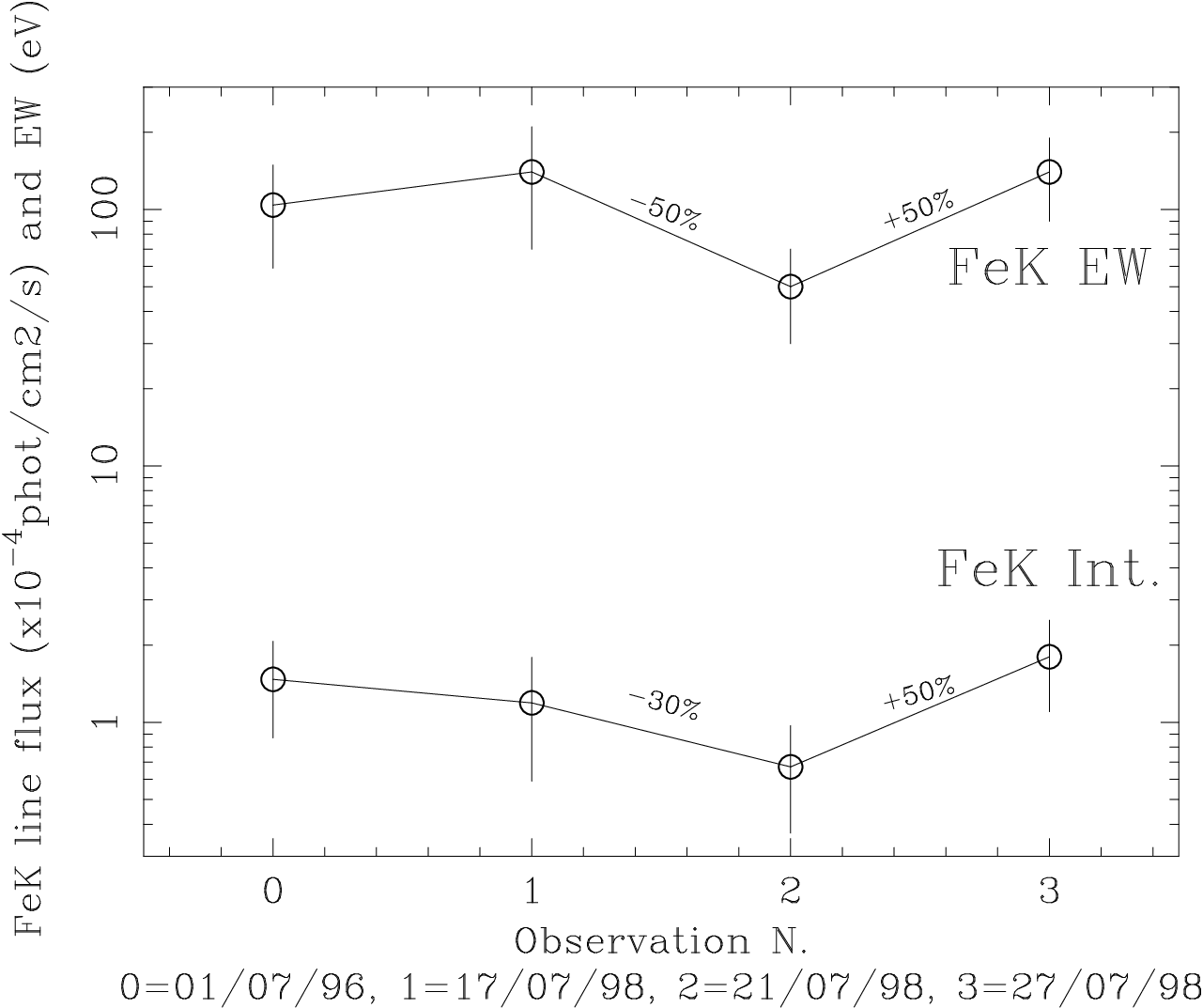,width=6cm,height=7cm,angle=-90}
\caption[h]{\small FeK line EW (upper points) and Intensity (lower points)
during Obs. 0,1,2 and 3}}
\parbox{6truecm}
{\psfig{file=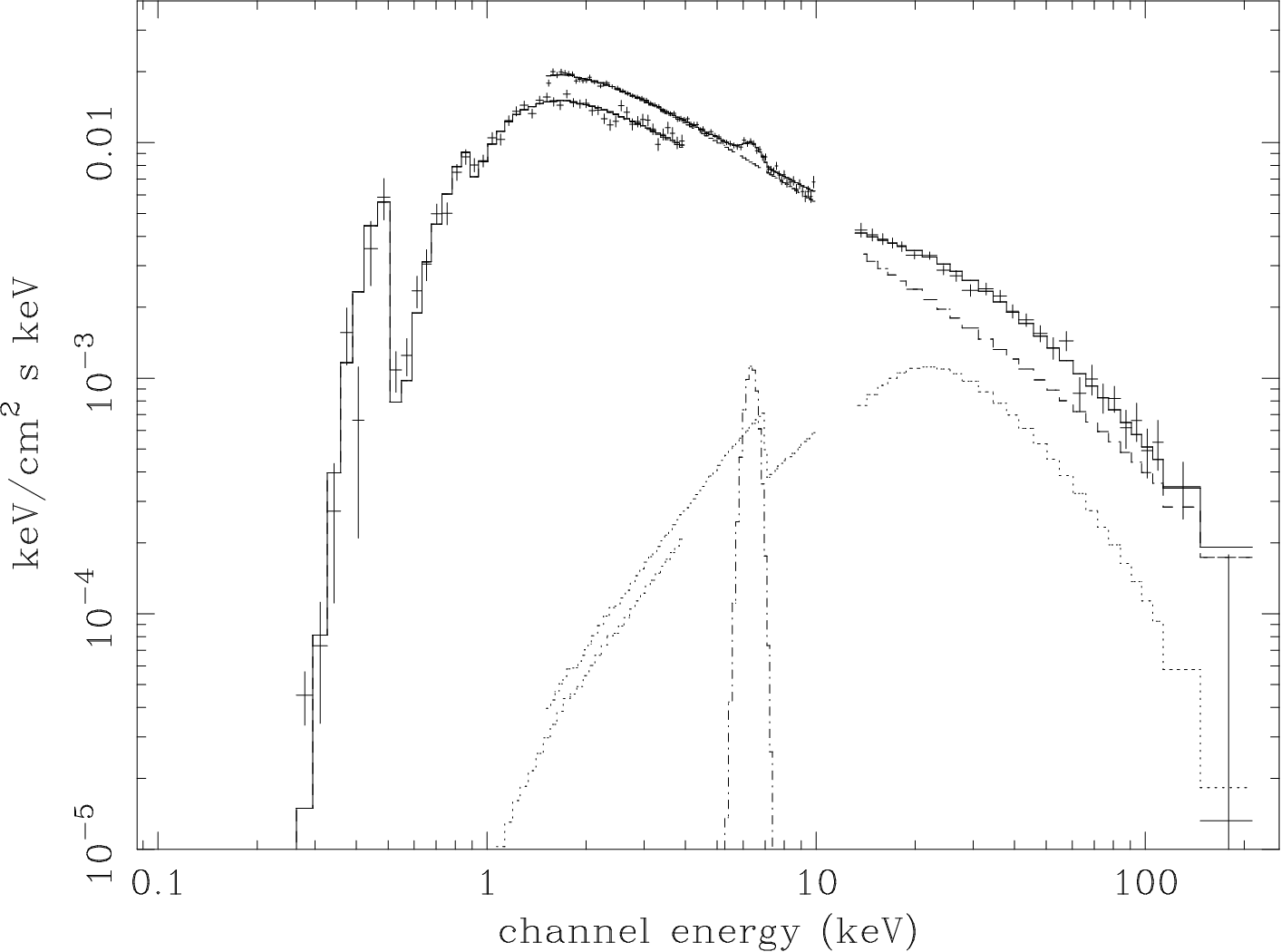,width=6cm,height=3.5cm,angle=-90}
\psfig{file=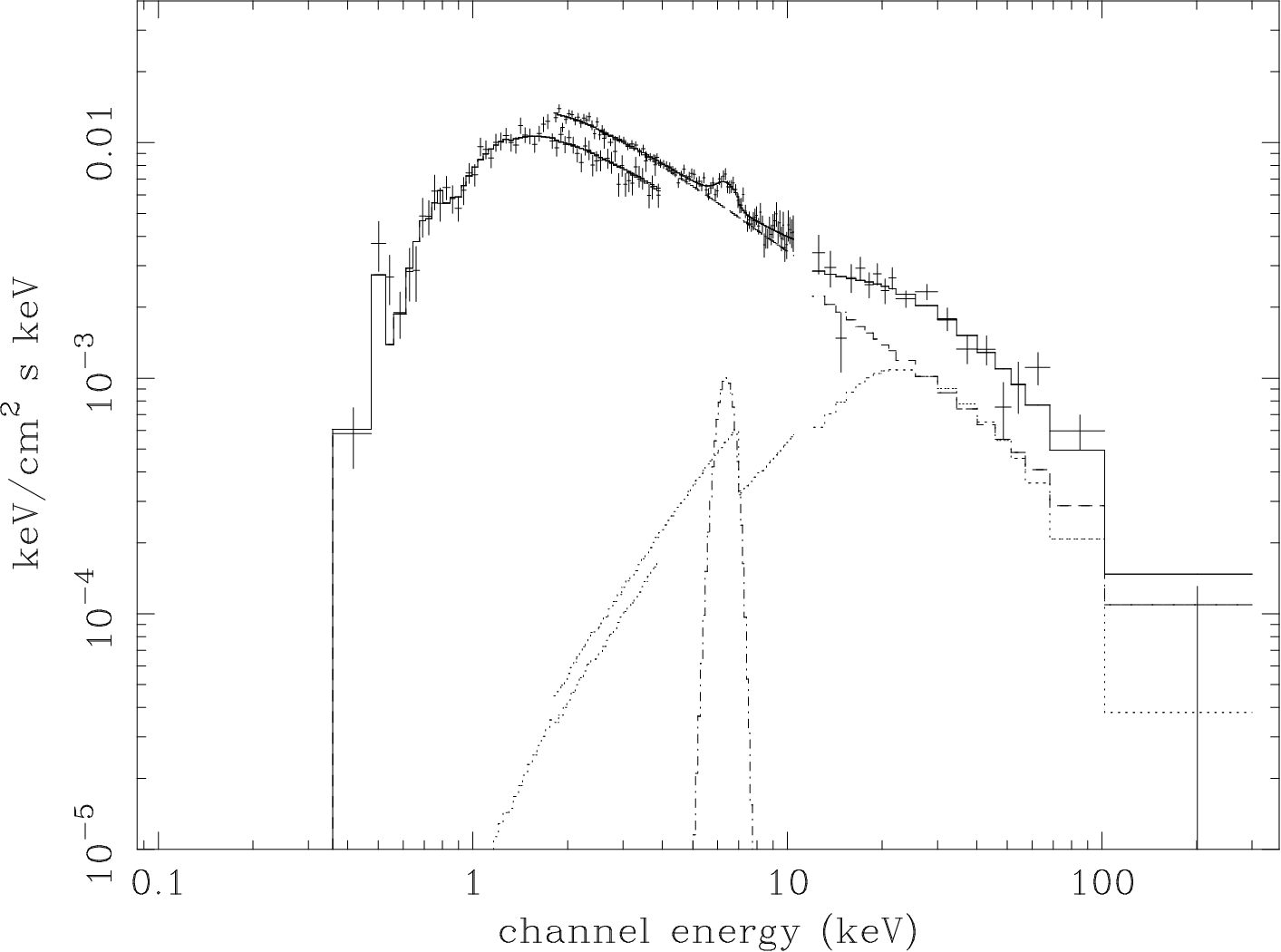,width=6cm,height=3.5cm,angle=-90}
\caption[h]{\small Best-fit unfolded spectra during the long 100 Ks
Obs. (top) and during Obs.1 (bottom) which illustrate the similarity of the spectra.}}
\end{figure}

\end{document}